\input epsf.sty
\input psfig.tex
\documentstyle[manuscript,eqsecnum,aps]{revtex}
\bibstyle{unsrt}
\begin{document}


\preprint{SUSX-TH-96-06 and IMPERIAL/TP/95-96/45}

\title{Out of Equilibrium Dynamics of a Quench-induced 
Phase Transition and Topological Defect Formation}
\author{Nuno D. Antunes$^1$ and Lu\'{\i}s M. A.  Bettencourt$^2$}
\address{$^1$School of Mathematical and Physical Sciences,
University of Sussex, Brighton BN1 9QH, U.K.}
\address{$^2$The Blackett Laboratory, Imperial College, London SW7 2BZ,
U. K.}
\date{\today}
\maketitle

\begin{abstract}
We study the full out-of-thermal-equilibrium dynamics 
of a relativistic classical scalar field through a symmetry breaking 
phase transition. In these circumstances we determine the evolution of
the ensemble averages of the correlation length and 
topological defect densities. This clarifies many aspects of the
non-perturbative dynamics of fields in symmetry breaking phase
transitions and allows us to comment on a quantitative basis on the
canonical pictures for topological defect formation and evolution.
We also compare
these results to those obtained from the field evolution in
the Hartree approximation or using the linearized theory. By doing so we
conclude about the regimes of validity of these approximations.

\end{abstract}

\vskip2pc


\newpage

\section{Introduction}
\label{s1}

The formation of topological defects  is a general
consequence of symmetry breaking phase transitions in field theories with a
topologically non-trivial vacuum manifold, both in the early Universe
\cite{Kib,Book}, 
and in a number of materials in the laboratory 
\cite{ZurekReview,Helium4,Helium3,liqcrys}. 

To date predictions of the number and distributions of defects formed
in these circumstances has relied on very simplistic heuristic models
where some qualitative aspects of the phase transition are invoked but
where all dynamics is  sacrificed \cite{VV}. These are at the basis of large
scale simulations of defect networks subsequently used to generate the
energy density perturbations responsible for the formation of structure in
the Universe \cite{Struc}.

Recently, considerable effort has been devoted to the development of
more realistic methods to follow the approximate evolution of
relativistic field
theories in out of equilibrium settings \cite{quench,non-eq,LA} and, in 
this context, to account for the number of defects  formed \cite{Ray}
at a symmetry breaking transition.

In this paper we perform the first complete fully non-linear 
dynamical study of a relativistic classical theory  out of thermal 
equilibrium in a symmetry breaking phase transition. We compute the
time evolution of many quantities of interest, such as the 
correlation length and the defect densities  as well
as their dependence on the choice of initial conditions and on the
presence of external dissipation.

We then compare these results to other approaches found recently in
the literature. We will show by explicit computation of
the time evolution for the zero densities of our field in well
specified and illustrative circumstances that both the Hartree
approximation and the linearized theory have different merits in
approximating the full classical evolution. By virtue of this
comparison we also learn, in a quantitave manner, when they fail and
consequently about the regimes of their applicability.

The results of this paper raise many extremely interesting new questions 
concerning the non-perturbative dynamics of relativistic fields 
away from thermal equilibrium. 
Our present methods rely heavily on the usage of extensive
computational facilities. 
Rather than openly trying to tackle some of these new issues  
the intention of the presentation below will  be a more 
modest one, openly restricted in its  character to that of reporting on
the finds of a numerical experiment. 
A more analytical approach must be
sought in order to complete our understanding, though,  and it is 
our intention to expand on our attempts in forthcoming publications.
Nevertheless, we believe the present results constitute 
considerable quantitave progress on the usual 
canonical qualitative pictures of defect formation and evolution.
 
This paper is organized as follows.
In section \ref{s2} we describe the theoretical background for the
field evolution. We  discuss the field equations for the full classical
evolution, in the presence of external dissipation, 
and our choice of initial conditions, which for consistency
we choose to follow a classical Boltzmann distribution. We then
describe two approximation schemes to the classical evolution, namely
the Hartree approximation and the linearized theory. In the latter
case we present the exact analytical evolution and 
Halperin's formula for the zero densities  of the scalar field
in a Gaussian theory. We finish this section by briefly describing
our numerical procedure for the full classical  evolution and the
computation of approximate statistical ensemble-averaged quantities.   

In section \ref{s3} we present our results. We show the
fully non-equilibrium evolution of the scalar field correlation length 
as well as its zero and defect densities. We show explicitly that the latter  
can be counted  at  a  given sensible field coarse-graining
scale, independently of the  necessary ultra-violet cut-off of our
implementation. We also discuss the dependence of these results on our
choice of initial conditions.
We then proceed to compare these zero densities to those obtained from the
Hartree approximation and the linearized theory, and to draw conclusions
about the regimes of their applicability.
Finally we present the evolution of the defect densities per
correlation volume and seek to relate our results to the qualitative
canonical arguments for defect formation and
long time evolution: the Kibble mechanism and scaling conjectures.

In section \ref{s4} we summarize our most important results, 
present our conclusions and point to questions raised by the present
work we intend to study in the near future.

\section{Theoretical Background}
\label{s2}

In what follows we will be concerned solely with the evolution of a
{\em classical} relativistic field theory. The role of quantum fluctuations
on the evolution of our field is therefore simply neglected. 
This is done in the spirit of Statistical Relativistic Field
Theories, e.g., \cite{Parisi}.

Our neglect of all quantum aspects 
should, however, constitute an excellent approximation to the
complete field dynamics at energies close to the phase transition
temperature. This is a non-trivial statement to quantify in general 
but that can be
made precise by treating the presence of the non-linear term in the
usual perturbative loop expansion. The 
fundamental difference between the thermal and the quantum 
loop perturbative series then arises from the fact that the quantum expansion
is organized in terms of increasing powers of $\hbar$ while the thermal
one is proportional to powers of the temperature $T=1/\beta$, made
adimensional by explicitly appearing in ratios with the mass scales
in the theory. 
In the regime in which the
temperature to mass ratios are 
very large compared to $\hbar$, quantum fluctuations
can be safely neglected relative to their thermal equivalent. 
Another manifestation of the quantum nature of a field theory is that
the energy spectrum becomes discrete and  consequently 
the thermal distribution follows a Bose-Einstein distribution 
instead of its classical Boltzmann form. Both classical and quantum 
distributions coincide approximately for frequencies $\omega$ such that 
$\omega/T << 1$, showing again that  quantum manifestations   
should become fundamental in the ultraviolet of
the theory and/or at low temperatures, as is well known. Both these regimes
are relatively unimportant for the studies presented below 
and, given our choice of the initial field configurations, may be probed only
under extreme conditions in the evolution or measurements performed
over the deep ultraviolet sector of the theory. 
We will not be concerned with these regimes below and that, 
given the fundamental difficulties inherent to a full
quantum approach relative to the opportunities offered by the
classical theory, constitutes our best justification for neglecting
quantum fluctuations. 
Although somewhat unsatisfactory we believe this is in itself
justifiable in view of the extremely interesting possibilities 
it permits, especially in opening a window for probing non-perturbative
aspects of the field evolution at the phase transition. 
In this section we will proceed to describe the details of
our full classical evolution as well as the basis for two
approximations, the Hartree self-consistent evolution and the
linearized theory. We conclude this section by presenting the
predictions from the exact results possible in Gaussian theories and
discussing our numerical methods.  

\subsection{The Classical Theory}
\label{ss2.1}

In what follows we will adopt the simplest
collisional  model displaying topological defects, i.e., we will be
dealing with a classical $\lambda \phi^4$ scalar field theory 
in $1+1$ dimensions. 
In one spatial dimension a Boltzmann distributed classical field is 
always finite and therefore does not require re-normalization \cite{Parisi}.
This choice of spatial dimensionality also allows us to  guarantee, from  
a technical point of view, that we will be able to evolve numerically  
a discretized dynamical system with enough resolution on all scales
and generate a sufficiently
large number of field realizations in order to be able to compute true
statistical ensemble-averaged quantities. In one  spatial
dimension and in equilibrium the Mermin-Wagner theorem \cite{MW}
states that there is no long range order. In this sense there is no
phase transition as understood canonically in terms of
Thermodynamic quantities. In an out-of equilibrium evolution such
as ours, however,  spontaneous symmetry breaking certainly occurs, in the
sense that the field chooses locally in space to fall towards either of the
energetically equivalent vacua. In what follows we will therefore
continue to use the term symmetry breaking phase transition in this sense. 
In any case, equilibrium in our evolution, as will
be clear by the end of this section, can only  be strictly 
achieved at zero temperature, where the Mermin-Wagner
theorem ceases to apply.
Field Theories in higher spatial dimensions will be 
considered in forthcoming  work \cite{ABY}.

In order to trigger the transition  
we set up initially a large number of field configurations
out of a micro-canonical statistical ensemble and, at $t=0$, 
destabilize the system by changing instantaneously the sign and
magnitude of the mass. 

The evolution equations for $t>0$, will then be taken to be 
\begin{equation}
\left( \partial_{t}^2 -\nabla^2 \right) \phi - m^2 \phi
+ \lambda \phi^3 +  \eta \dot{\phi} = 0,
\label{e2.1.1}
\end{equation}
where $\lambda$ is the scalar self-coupling, and $m$ the classical
mass. The dissipation coefficient $\eta$ is included as we wish to
describe a system in the presence of additional degrees of
freedom. This a necessary condition in order to justify the use of a 
micro-canonical thermal distribution of fields as our 
initial conditions.
Accordingly, the form of the evolution equations can be exactly
obtained if our system, the scalar field, is assumed to be in contact
with a much larger one, a  thermal
bath of oscillators, linearly coupled to $\phi$.
The result is a  Langevin system with a simple Markovian dissipation
kernel, characterized only by the constant value of $\eta$, 
and a noise term, 
where the two are related by the fluctuation-dissipation theorem 
\cite{Kabib}. 
At zero bath temperature the system reduces to Eq.~(\ref{e2.1.1}). 

It is convenient and physically clarifying to consider the
rescaled system  by redefining the time and space variables
as well as the field amplitudes as in
\begin{equation}
x \rightarrow x/m, \qquad t \rightarrow t/m, \qquad \phi \rightarrow
 \sqrt{\lambda}/m ~ \phi,
\label{e2.1.2}
\end{equation}
to be
\begin{equation}
\left( \partial_{t}^2 -\nabla^2 \right) \phi -  \phi
+  \phi^3 +   \tilde{\eta} \dot{\phi} = 0,
\label{e2.1.3}
\end{equation}
where 
\begin{equation}
\tilde{\eta} =  \eta/m. 
\label{e2.1.4}
\end{equation}

The new dissipation constant, $\tilde{\eta}$, has a clear physical  
interpretation. It is the ratio of the wave-length $1/m$, 
for the mean scalar field to the diffusion length, $1/\eta$, 
of a free Brownian particle  in contact with the thermal bath.
Whenever this ratio is small the diffusion length is large compared to
the particle's wave length and the only collisional effects present
involve the interaction of the scalar field with itself. In the
converse limit the system is dominated by dissipative effects to the
bath and the dynamics obeys essentially a diffusion equation
where the second derivative in Eq. (\ref{e2.1.3}) is negligible
relative to the first and as a consequence
oscillations and self-interactions  play a secondary role.

The ratio $\tilde{\eta}$ 
is therefore our fundamental dynamical parameter. All other
information about the system is encoded in the initial field
configurations. 
To specify these we  assume that, for $t<0$, the system  
will be described by a free-field equation, with possibly a different 
mass parameter, $M$,
\begin{equation}
\left( \partial_{t}^2 -\nabla^2 \right) \phi + M^2 \phi = 0,
\label{e2.1.5}
\end{equation}
in thermal equilibrium at a given temperature $1/\beta$.
Our goal is to compute ensemble averages of several quantities
throughout the evolution.
In order to achieve this we generate initially a large number of field
configurations out of a Boltzmann distributed statistical ensemble. 
This is, off course, the classical micro-canonical equilibrium distribution.
The probability density functional, $P[\phi,\Pi]$, for the field and its 
canonical conjugate momentum $\Pi= {\partial \phi \over \partial t}$ will
be given by
\begin{equation}
P[\phi,\Pi]\propto e^{- \beta {\cal{H}}[ \phi, \Pi]},
\label{e2.1.6}
\end{equation}
where $\cal{H}$ is the free-field Hamiltonian
\begin{equation}
{\cal{H}}[ \phi, \Pi] = {1 \over 2} \int_{0}^{L}dx \; \Pi^{2}(x) + 
\left( {\partial \phi\over\partial x}(x)\right)^{2} + M^{2} \phi^{2}(x),
\label{e2.1.7}
\end{equation} 
where $L$ is 1-dimensional volume of our system, which will be taken to be 
much larger than the mean correlation length of the initial field
configuration $\xi=1/m$.

Since this is a Gaussian distribution we need only specify the 
mean-value and variance for the field and its conjugate 
momentum in order to characterize it.
We do this most simply in Fourier space, where we have  
\begin{equation}
<\phi_k(0)> = 0, \quad  \quad 
<\phi_k(0) \phi_{-k}(0)>  =  {1 \over \beta L w_k^2}
\label{e2.1.8}
\end{equation}
and
\begin{equation}
<\Pi_k(0)> = 0 \quad  \quad
<\Pi_k(0) \Pi_{-k}(0)>  =  {1 \over \beta L}, 
\label{e2.1.9}
\end{equation}
with the dispersion relation
\begin{equation}
\omega_{k}=\sqrt{k^{2} + M^{2}},\quad k={2 \pi n \over L},
\quad n=0,1,...,+\infty. 
\label{e2.1.10}
\end{equation}

The same statistical field configuration  could have  been obtained by
driving a linear (i.e., free of self-interactions) field to thermal 
equilibrium in contact with a  reservoir at 
temperature $1/\beta$, using a Langevin evolution
equation, \cite{Parisi}.

In order to trigger the symmetry breaking transition we proceed,  
at $t=0$, to instantaneously  change $M^2$, in sign and in magnitude, 
thus forcing the system
to leave thermal equilibrium and to evolve in a non-linear way,
according to Eq.~(\ref{e2.1.1}). This is the simplest way of de-stabilising
the system and has the advantage of analytical tractability in the
simplest cases. Because of this  property it has been
extensively used in the literature \cite{quench,Ray}, 
where the
instantaneous change of $M$ is often referred to as a {\em quench}.
In terms of bulk thermodynamical quantities it corresponds to a sudden
decrease in Pressure, or in the language of Finite Temperature Field
Theory a decrease in the theory's effective potential. 
It is not clear to us, however, if such a triggering mechanism  has
a natural implementation in the Laboratory \footnote{It is clear that
this is not the mechanism of destabilising  the theory in a
cosmological context. Pressure quenches were used to drive liquid
$^4$He systems through a  super-fluid phase transition \cite{Helium4} 
but are thought to be accompanied by other energy loss mechanisms.}.

In order to compute average quantities  we then evolve numerically 
a large number of random field realizations out of the statistical ensemble 
Eq.~(\ref{e2.1.8}-\ref{e2.1.9}), using Eq.~({\ref{e2.1.1}) 
and compute, at given time-intervals,  their mean values
over this set.

Before we proceed to present our results we will briefly describe two
approximate schemes to the evolution described above, the Hartree
approximation and the linearized theory. In the next section we
will compare the results obtained by these three different approaches.  

\subsection{The Hartree Approximation}
\label{ss2.2}

A widely used approximation scheme to the full  approach
described in the previous section  is the so called Hartree 
self-consistent approximation \cite{HF}. 
Unlike the naive perturbative expansion it has the virtue of remaining  
stable throughout the symmetry breaking transition. It is also  
fully renormalizable in the quantum case \cite{non-eq,LA}.
Because of these characteristics the Hartree approximation has  
received quite a lot of attention in  the recent
literature as a means of performing out-of equilibrium computations in
various situations \cite{non-eq,LA}.

As a drawback it describes a theory where all energy transfers must be
made through the mean-field and, as a consequence, in situations
when the field evolution involves important energy transfers among
modes with $k \neq 0$,  it behaves poorly.
Our objective below will be to make the latter statement more precise.    

The Hartree approximation  results from making the Lagrangian 
quadratic, by replacing the quartic term by an expression involving
the average value of the quadratic field only. 
More interestingly,  it can be seen to arise  as the first order 
in a systematic  perturbative expansion,  in the parameter $1/N$ of
an $O(N)$ symmetric theory, see, e.g., \cite{LA}.

Once this form for the Lagrangian density is assumed the corresponding
evolution equations  are required to be self-consistent, in the sense
that the  zero separation two-point function computed at each step  
of the evolution is the same one present in the  
effective Lagrangian, for the corresponding time.

In practice  this can be achieved by  replacing the cubic term in the 
Euler-Lagrange equation by 
\begin{equation}
\phi^3 = 3 <\phi^2> \phi.
\label{e2.2.1}
\end{equation}
With this substitution the approximate evolution equations now 
describe a Gaussian field with an effective time-dependent mass. 
This allows us  to easily compute the time dependent two-point function 
$W(x,x',t,t')$,
\begin{eqnarray}
W(x,x',t,t')&=&\langle\phi(x,t)\phi(x',t')\rangle \nonumber \\
&=& \sum_{k=-\infty}^{+\infty} G_{k}(t,t') e^{i k (x-x')},
\label{e2.2.2}
\end{eqnarray}
where we chose to write $G_{k}(t,t')$ in terms of the usual positive
and negative frequency  modes,
$U^{+}_{k}(t)$ and $U^{-}_{k}(t)$ respectively, as 
\begin{equation}
G_{k}(t,t') =  U^{+}_{k}(t) U^{-}_{k}(t')+U^{-}_{k}(t) U^{+}_{k}(t').
\label{e2.2.3}
\end{equation}

These, in turn,  obey the field evolution equations, 
\begin{equation}
\left[ {d^2 \over dt^2} + k^2 - m^2 + 3 \lambda
<\phi^2(t)>  + \eta {d \over dt}\right] U^{\pm}_n(t) = 0.
\label{e2.2.4}
\end{equation}

The initial conditions Eq.~(\ref{e2.1.8}) and (\ref{e2.1.9}), 
can also be written in terms of the positive and negative frequency 
modes  $ U^{+}_{k}(t)$ and $ U^{-}_{k}(t)$. We obtain
\begin{equation}
U^{\pm}_n (t=0) = {1 \over  \omega_n \sqrt{2 \beta V}}; \qquad 
\dot U^{\pm}_k (t=0) = {\pm i \over \sqrt{2 \beta V}}.
\label{e2.2.5}
\end{equation}
For the Hartree scheme to be complete we  have to impose a 
consistency equation, namely that the mean-field taken in Eq.~(\ref{e2.2.4})
be the same as the result obtained from two-point correlation function
calculated from  Eq.~(\ref{e2.2.3}). This gives,
\begin{equation}
<\phi^2(t)> = 2 \sum_{k=-\infty}^{+\infty} U^{+}_k(t) U^{-}_k(t).
\label{e2.2.6}
\end{equation}

Equations~(\ref{e2.2.4}) and (\ref{e2.2.6}) together with
the initial conditions Eq.~(\ref{e2.2.5}) constitute 
a well posed initial value problem 
that can be easily solved numerically, for a discretized set of
momenta.  Note that once we obtain the 
time-dependent correlation function we have at our disposal all the information
about the system since the Hartree approximation implicitly assumes
a Gaussian distribution of fields.

Eq.~(\ref{e2.2.4}) shows that the 
Hartree approximation is clearly  collisionless as it describes the
interactions of an infinite  set of modes with a mean field. As a
result energy transfer processes among the modes 
proceed without any exchange of momentum, necessarily through the
mean field. 
The collisionless  character of the Hartree approximation
constitutes its greatest weakness and
will bring about substantial differences  to the behavior observed by
evolving the scalar field using 
Eq.~(\ref{e2.1.1}), as we will illustrate in the next section.

\subsection{The Linear Approximation}
\label{ss2.3}

More severely one can neglect the interactions altogether simply by
removing the cubic term in the evolution from Eq.~(\ref{e2.1.1}). This is
necessarily an extremely crude approximation but has the merit of
making it possible to predict, given an initial Gaussian field
configuration, all the relevant quantities analytically. It can be
also assumed [as it has been done frequently in the literature, see
\cite{Ray}] that for certain parameter ranges 
the relevant evolution occurs in the period soon after the field 
leaves equilibrium and starts descending towards the minimum of the
new potential. During this stage the potential can, in fact, be approximated 
by an inverted squared well, but as we will see later the non-linear 
aspects of the evolution will in well defined circumstances 
be  relevant for the mechanism of defect production, altering
substantially their numbers.

 Since the evolution is linear, the distribution of fields will remain
Gaussian for all times and we need only  determine the two-point 
correlation function in order to have a full description of the
system. This can easily be done analytically, with the positive and
negative frequency modes being given by
\begin{eqnarray}
  U_{k}^{+}(t)&=&{e^{-\eta/2\;t} \over\sqrt{2\beta L}}
 \left({1\over \omega_{k}}\; \cosh(\Omega_{k}
  t) \right. \nonumber \\
&+& \left.{i\over \Omega_{k}}(1 - {i \eta\over 2 \omega_{k}})
\;\sinh(\Omega_{k} t)\right)\;\;\;\vert k \vert < k_{c}
\label{e2.3.1a} \\
  U_{k}^{+}(t)&=&{e^{-\eta/2\;t}\over\sqrt{2\beta L}}
  \left({1\over \omega_{k}}\;
 \cos(\Omega_{k}
  t) \right.  \nonumber \\
&+& \left.{i\over \Omega_{k}}(1 - {i \eta\over 2 \omega_{k}})
 \;\sin(\Omega_{k} t)\right)\;\;\;\vert k,
  \vert > k_{c} \label{e2.3.1b} \\
\end{eqnarray}
with
$$
U_{k}^{-}=(U_{k}^{+})^{*},
$$
where $\Omega_{k} =\sqrt{ \vert k^{2} - m^{2} - \eta^{2}/4\vert }$
 and $k_c  = \sqrt{ m^2 + {\eta^2 \over 4}}$.
Note that, modulo the effect of the external dissipation  that damps
all modes,  the  evolution is divided in the usual way between exponential
[growing for $\vert k \vert < m$ and decaying for 
$m<\vert k \vert < k_c$] and oscillatory  [for $ \vert k \vert > k_c$].

\subsection{Analytical Results for the Gaussian Theory - Counting
Zeros and Defects}
\label{ss2.4}

In the special case of the linearized evolution exact analytical
predictions are possible, which constitute an excellent test on our methods.
These results are  based on a well known computation, 
first derived by Halperin \cite{Halperin}. 
Given the knowledge of the equal time two-point 
function
and its spatial derivatives it allows us to calculate 
the average spatial density of zeros $\langle n_{0}\rangle$,
in a Gaussian distribution of fields. 
In one dimension Halperin's formula becomes,
\begin{equation}
 \langle n_{0}(t)\rangle = {1\over \pi}\sqrt {\left| {W''(0,t)\over 
W(0,t)}\right|},
\label{e2.4.1} 
\end{equation}
where, by translational invariance,
\begin{equation}
W(x,x',t)= W(\vert x-x' \vert,t)\equiv W(r,t).
\label{e2.4.2} 
\end{equation} 

One of the simplest application of this result is for the case of a Boltzmann
distribution of free quadratic fields, which corresponds to our initial
conditions. We then have that our two-point function is explicitly
time independent and  takes the form
\begin{equation}
   W(x,x')={1\over\beta L}\sum_{k=-\infty}^{+\infty}{1\over\omega_{k}^{2}}
   \;e^{i k (x-x')},
\end{equation}
on a discrete periodic lattice, 
with $k$ and $\omega$ obeying the dispersion relation, Eq.~(\ref{e2.1.10}).
When substituted in Eq.~(\ref{e2.4.1}) this gives
\begin{equation}
\langle n_{0}\rangle = {1\over \pi}\sqrt { \sum_{-\infty}^{\infty} 
   {k^{2}\over\omega_{k}^{2}} \over 
   \sum_{-\infty}^{\infty}{1\over\omega_{k}^{2}}}. 
\label{e2.4.3}
\end{equation}
It is clear that $\langle n_{0}\rangle$, given by  Eq.~(\ref{e2.4.3}),
diverges. Introducing an upper momentum cutoff $\Lambda$ we have that this
 is of the form  $\langle n_{0} \rangle \sim \Lambda^{1/2} $ 
\footnote{For spatial dimension $D=2$, $ \langle n_{0}\rangle \sim
\Lambda^{2}/\log(\Lambda)$. For $D \geq 3$, $\langle n_{0}\rangle 
\sim \Lambda^{D}$.}.

Eq.~(\ref{e2.4.1}) can be easily extended to both the Hartree and the
linear evolutions, since in both cases the field distributions
remain Gaussian for all times. This fact allows us to 
obtain a time dependent zero density, which can be written explicitly
in terms of the propagating modes as
\begin{equation}
 \langle n_{0}(t)\rangle =  {1\over \pi}\sqrt {
   \sum_{k=-\infty}^{+\infty}\; k^{2}\;\vert U_{k}^{+}(t) \vert^{2} \over
   \sum_{k=-\infty}^{+\infty}\vert U_{k}^{+}(t) \vert^{2} }.
\label{e2.4.4}   
\end{equation}
As before the result clearly diverges. This can be seen from the explicit 
analytical result, in the linear case, or numerically for the Hartree
approximation.
This divergence
is  a consequence of the existence of a too large number of zeros in
arbitrarily small scales, too large in fact for the sums in 
Eq.~(\ref{e2.4.4}) to converge to a  finite result. 
The number of zeros that correspond to
defects is, however, clearly independent  of the behavior of the 
field deep in the ultra-violet. 
In order to measure the correct number of defects present in a given
field configuration with zero crossings on all scales 
we must therefore introduced a 
coarse-graining scale, which we will naturally chose to be of the order
[usually slightly larger] than
the size of a defect \cite{Habib}. In one spatial dimension  there is an exact
domain wall solution with a well defined width, given by $1/m$. 

When evolving the full non-linear theory  we will thus have to introduce 
two relevant scales. An upper momentum cutoff, $\Lambda$, which is
related to
the  number of modes chosen to generate the initial conditions, 
and a coarse-graining
scale, of the order of $1/m$, used for calculating the defect density. 
This density has then to be shown explicitly
not depend on the upper momentum cut-off. Given this cut-off scale 
we will also measure
the zero density and compare it to the analytical predictions both 
from the linear and Hartree approximations.
 
\subsection{The Numerical Evolution}
\label{ss2.5}

In order to perform the non-linear out-of-equilibrium evolution we
have used a 128-processor parallel computer.
We took advantage of this architecture to evolve in each processor a
different random realization of the initial Boltzmann distributed
scalar field. For  each  one of these 
we took the initial ultraviolet cut-off to correspond
to a wave number of about 1000 and  our spatial lattice to
have 10625 sites. A defect's width was always resolved with more than 12
lattice points.  
All quantities of interest were then computed at given time intervals 
by averaging over the ensemble.

To perform the numerical evolution we used a Second Order Staggered
Leapfrog  method. The corresponding set of equations for the field and
its conjugate momentum are   
\begin{eqnarray}
&\Pi& (x,t+1/2 \delta \;t) =\frac{1-\chi}{1+\chi}\;\Pi(x,t-1/2 \;\delta t) +
 \nonumber \\
 &\;&\;\;+ {\delta t \over 1+\chi} \; \,
 [ \nabla^{2} \phi(x,t) + \phi(x,t) - \phi^{3}(x,t)]  \\
&\phi&(x,t+\delta t) = \phi(x,t) + \delta t \; \Pi (x,t+1/2 \delta \;t),
\label{e2.5.1}
\end{eqnarray}
where $\chi = \tilde{\eta}\;\delta t /2$ and $\delta t$ is the time step.

The initial conditions were generated in Fourier space using a Normal
Distributed random number generator and then converted to real space
using a Fast Fourier Transform algorithm.
For each chosen time step  we measured the average density of zeros
of the field [ by looking at sign changes at consecutive lattice points], 
the average density of defects [by counting
the number of zeros in the coarse-grained field ] 
and the correlation function.  
Using the correlation function we have calculated the correlation length
which we defined as the point at which the value of the 
correlation function, normalized at
zero spacing to be unity, goes below $1/e$. This also enables us   
to define defect and zero densities per correlation length volume.

Several precautions should be taken in order to guarantee 
a good accuracy of the results. The spatial step should be small
enough to resolve the defects  
and  the time step should obey the Courant condition
$$
\delta t << \delta x,
$$
where $ \delta x$ is the physical lattice spacing, 
in order for the method to converge safely.  

\section{The Results}
\label{s3}

\subsection{Testing our methods and exact analytical results for the
Gaussian Theory}
\label{ss3.1}

The simplest test we can perform on our procedure is to measure the
zero densities in our numerical evolution in the special case of
$\lambda=0$ and compare the  results
to the exact predictions of the linearized theory.

This involves several aspects of our numerical data.
Firstly we want to test whether our randomly 
generated initial conditions reproduce
faithfully an average field configuration out of  Boltzmann distribution 
and, in the affirmative case, whether the numerical evolution in the
simple case of the linear theory coincide with the exact analytical
results derived in sections \ref{ss2.3} and \ref{ss2.4}.

In order to do the latter we must  necessarily introduce
an ultraviolet momentum cut-off,  $\Lambda$. 
We note, however, that Halperin's formula is still applicable   
for a finite number of modes,  the exact result being 
given by  Eq.~(\ref{e2.4.4}) with finite limits in the sum.
Therefore, the unavoidable introduction of a cut-off 
should not be an  obstacle to the numerical  verification of the 
analytical predictions.
Before  comparing the zero density obtained from the numerical 
linear evolution to the exact result we must pay attention to 
one last technical complication.
When converting a finite set of amplitudes in $k$-space to $x$-space 
using the Fast Fourier
Transform algorithm we are actually loosing resolution, in the sense
of field structure, in the smallest scales, which in turn can lead to an
underestimate of the number of zeros. 
We explicitly observed this problem when transforming a generated
field configuration with $N$ amplitudes in momentum space to a grid
with the same number of points in configuration space. 
A simple way of overcoming this difficulty 
is to, for a chosen  $\Lambda$, generate an extra number of modes of
higher momentum with zero amplitude and then transform this set to $x$-space.
This corresponds basically to an increase in the number of points in real 
space (and thus in the spatial resolution) while keeping a given momentum
cut-off fixed. How much precision we actually need is then  
decided by increasing the number of extra modes with zero amplitude
until the average zero density converges
to the analytical result for the Boltzmann distribution, given by 
Eq.~(\ref{e2.4.4}). We were careful to follow this procedure even when the
comparison to the linear results was unnecessary, in the case of the
full classical evolution.

Having done this, we obtained an excellent agreement between the 
initial conditions and their numerical evolution and the exact analytical 
result, with a very small standard deviation
within our ensemble, represented by the error bars in  figure \ref{fig1}.

This also reassured us that the number of samples used in our
ensemble, 128 as mentioned above, is large enough for us  to 
obtain a good approximation to the exact ensemble averaged quantities.

\subsection{Non-linear Field and Defect evolution}
\label{ss3.2}

Having performed the tests of subsection \ref{ss3.1}, we are 
now ready to analyse the evolution of the field in the presence
of the non-linear term. As  mentioned before, the evolution results
depend only on one dynamical parameter $\tilde{\eta}$ and on the initial
conditions which are completely specified by the values of the
temperature, $1/\beta$, and the mass, $M$. In what follows we will
assume the re-scaled evolution of Eq.~(\ref{e2.1.3}) and drop tildes.

An example of the zero and defect density evolution is shown
in fig. \ref{fig2}. 
As expected the density of the latter is
always smaller than that of the former and independent of the chosen
momentum cut-off. We also observe that for reasonably dissipated
systems the defect and zero densities coincide for large times.
The zero densities observed in the coarse-grained Gaussian initial
field should not be taken strictly as defects, off course, as they lack the
stability only obtained at later times as the field truly settles down
locally at either of the energetically equivalent vacua.

The qualitative evolution of the field, in the aftermath of a sudden 
quench,  can be seen to follow in general terms two quite  
different stages. 
Firstly, immediately after the quench, the
negative curvature of the potential near the origin gives rise to
instabilities in the fields, in the sense that if one neglects the
cubic term in the evolution, which is initially taken to be small
[because $\langle \phi^2(t=0)\rangle << 1$, by construction] 
the modes with momentum $k < k_c$ will evolve according to
the real exponential forms of Eq.~(\ref{e2.3.1a}-\ref{e2.3.1b}).
The corresponding exponential growth 
distorts the original field configuration since the amplitudes for
the unstable modes grow much larger than their  characteristic
value, typical of  the thermal initial conditions, while the remaining
amplitudes stay approximately the same, but for damping if 
dissipation is present. This represents a considerable deviation from
the Boltzmann distributed initial field. 
This characteristic unstable behavior 
suggests that it is a good working hypothesis to assume  the field in this
initial stage  follow approximately the linearized evolution. 
We will check and confirm this conjecture in what follows.

The instabilities shut down when the cubic term, in Eq.~(\ref{e2.1.3}),
grows large enough to compensate for the negative sign of the mass. 
This is the beginning of the second stage of the evolution, 
often referred 
to in the literature as re-heating since the field  during this
period  tends again to a new 
maximum entropy configuration. The evolution at this stage is
completely non-linear and inaccessible through the usage of the usual
perturbative loop expansion. 

At the beginning of re-heating  the fields are severely red-shifted relative 
to a thermal distribution and energy
redistribution among the modes must occur. As a result, in the absence
of strong dissipation, 
the amplitude of the short wave-length modes grows while that of the
long wave-length modes decreases. These two types of behavior are now
not just  characteristic of modes with momentum 
$k>k_c$ and $k<k_c$, respectively, since among the latter the
amplitude has also grown differentially, approximately 
with the exponential of their characteristic frequency
Eq.~(\ref{e2.3.1a}).
As a result of this flow of energy to smaller wave-lengths the field
configurations change to display much more structure on
smaller scales including those in which topological defects can be
produced.
After the first burst of energy transfer to smaller scales the
resulting field configurations involving large gradients in small 
scales are
energetically unfavoured and the field quickly evolves back to 
suppress them partially. This results in a series of 
oscillations in typical
quantities, such as the zero densities  or the
correlation length, fig. \ref{fig3} and \ref{fig8}.

In the presence of external dissipation, for $\eta$ non-zero, the
instabilities grow in an analogous way, soon after the quench, but the 
process of energy redistribution or re-heating 
can be quite different. The presence of external dissipation, as we
discussed above, can be seen  as resulting from the existence of effective
channels, i.e., modes of other fields,  
which compete with those of the scalar
field for the energy transfered from its largest wave-length modes.
The value of $\eta$ then determines the relative importance of these
two types of channels. This competition among scalar field and
channels external to it  turns out to be absolutely crucial 
for the form of the evolution of the zero and defect densities.

Schematically, with our choice of parameters, for $\eta=1$,
the system is always strongly dissipated and behaves much like a field
without self-interactions [albeit stable] that freezes in very
early. This can be clearly seen in the field profiles of fig. \ref{fig5}. 
The effective external
channels therefore strongly dominate over the scalar field self-
interactions and no signature of re-heating, such as the creation of
zeros or a strong drop in correlation length, is observable.

For smaller values of the dissipation kernel $\eta=0.1-0.05$, both
self-interactions and dissipation are important, acting on the same 
kind of time scales. 
It is clear from our results that the transfer of amplitude among the
modes occurs at well-defined stages of their oscillations. This is a
subject of study in its own right, which is beginning to receive  much
attention in the context of the theory of re-heating after a period of
inflationary expansion in the early Universe. Our objective in this
paper is not to tackle this question but, in view of the strong
analogies between the two problems, to merely  point out that as
a consequence of this behavior the creation of zeros and defects 
at re-heating proceeds by bursts. 
This is visible, e.g. in fig. \ref{fig4}.

When zeros and defects are created in these bursts the dissipation 
[ if not high enough ] and the field self-interactions are not
sufficiently effective to suppress them immediately. 
As the momentary production shuts off, however, 
both these processes reduce the zero densities considerably. 
At the next burst another large amount of small scale 
structure can again be  created but 
smaller than that at the  previous instance. 
The field evolution then
proceeds to dissipate it away and so on.
As a result of these two competing processes
the field oscillates between having quite large amounts of structure on
small scales to having little, as both processes of creation and
dissipation seem to be  most efficient after the converse one has acted.  
These processes are clearly visible in the  
large oscillations undergone by the correlation length and defect densities. 
Profiles out of a field evolution in this dissipative regime is shown
in fig. \ref{fig6}.

Finally, for small values of the dissipation coefficient, $\eta \leq 0.01$,
the field is allowed to reach a favourable configuration before the
dissipation has any sizable effect. 
The zero densities still oscillate in a similar fashion but field
ordering in small scales  now seems to be due to essentially
the action of the field self-interactions.  
For this same reason the densities for large times are much larger
that those obtained under the presence of larger dissipation.
Its effect, if present at all, is only to damp all 
modes for large times. Modes on smaller scales are more strongly
dissipated, though, due to the fact of possessing a larger 
natural frequency.  

Snapshots of the field out of a characteristic evolution 
under the effect of very weak dissipation can be seen in the field 
profiles of \ref{fig7}.

Equivalently, the field evolution can be studied qualitatively by
considering the time dependence of the 
correlation length. Fig. \ref{fig8}, shows the behavior of
the correlation length in time under the effect of several values of
the external dissipation.

It is clear that the correlation length increases in the first quasi-linear
stage of the evolution, decreases at re-heating and increases again as
the field organizes itself both via the effect of its
self-interactions and  due to the action of the external dissipation.

It is interesting to note, however, that the time evolution of the correlation
length for large times, after re-heating, seems to display  quantitatively
different time trends depending on whether 
the self-organization proceeds by essentially 
the action of the scalar field self-interactions or results from the
effect of the external dissipation. 
This seems to result in a linear time
dependence in the first case and a well-known diffusive behavior with
$t^{1/2}$ in the latter. This is illustrated in fig. \ref{fig47},
where it is also interesting to note that the slope of the linear
correlation length growth for small dissipation is about $0.1$ of the
speed of light.

This seems to suggest that the underlying  field ordering  proceeds by the  
free propagation of a well defined signal at a fraction of the speed 
of light as is often suggested in qualitative scenarios for domain
formation and  growth.

It is off course also possible that the linear behavior observed above
will constitute merely a transient regime and that the evolution of the
correlation length actually follows the same diffusive pattern,
obtained under the effect of larger dissipation, but on a much
larger time scale. For even smaller values of $\eta$ we observed the
persistence of the linear behavior, but with smaller slopes, showing
that the external dissipation is still playing a role in the field's
self-organization.

We also tried to determine the influence of our choice of 
initial conditions on the evolved defect densities.
Initially, for an average Boltzmann field in the continuum limit,  
the spatial two-point
correlation function  has the exact form
\begin{equation}
<\phi(x)\phi(x^\prime)> =  \pi {T^3 \over m} e^{-M \vert x-x^\prime
\vert}.
\end{equation}
This defines unambiguously the correlation length to be  $\xi=1/M$.
Different choices of initial conditions corresponding to different 
masses $M$ therefore lead to
quite distinct field configurations on given  scales.
Fig. \ref{fig9} shows the defect density evolution for a choice of
low dissipation, $\eta=0.01$, and three different initial masses. 

The initial temperature was chosen so as to guaranty that the field 
undergoes initial
instabilities, i.e, so as to ensure the effectiveness of the quench as the
mechanism driving the transition. The values chosen  were such
that $\langle \phi(t=0)^2\rangle> \sim 0.1$, in our original Boltzmann
distributed field. This results in small values of the temperature
relative to the mass scales, of the order $M/T \sim  0.01-0.05$. 
It should be noted however, that 
the initial zero and defect densities are  independent 
of this choice of temperature.

Given a choice of $M$ and the associated correlation length $\xi =
1/M$ we expect, in rough terms,  that 
the defect densities will  be larger for smaller $\xi$, as
this is a qualitative indication of more structure on smaller scales. 
This can be clearly seen by comparing the initial magnitudes of the defect
densities in figure \ref{fig9}, for three different choices of initial
correlation lengths.

Fig. \ref{fig9} shows a much more striking fact, however. It is clear
that the number of defects produced at the phase transition is
approximately independent of the initial densities even when these
differ by over an order of magnitude.   This is a clear indication of
the importance of accounting correctly for the number of defects
present at the time of re-heating and not sooner. This observed
independence of the choice of initial defect densities is only
characteristic of evolutions under the effect of small or null
dissipation as in strongly dissipated systems re-heating is,
as we have seen above, severely suppressed.
 
\subsection{Comparison with the results from the linearized theory and
the Hartree approximation}
\label{ss3.3}

In the previous section we presented and discussed the results for the
full classical evolution of Eq.~(\ref{e2.1.3}). 
In view of their extensive application to non-equilibrium problems in
the literature 
it is extremely interesting to analyse how particular approximations
to this classical theory perform relative to it.

\noindent In this subsection we briefly compare the results for the zero
density evolution given by  the full classical evolution 
to those obtained  in the  Hartree evolution and 
by considering the free field given by the  linearized theory.

Fig. \ref{fig11} and \ref{fig12} show two examples of the  zero 
density evolution computed in these three cases, and for relatively low 
and high dissipation, respectively.
The observed difference between the zero densities given by the 
full classical evolution and those of the 
linear theory is quite simple to understand. 
For large $\eta$, as we discussed above, the effective channels
present implicitly in the form of the dissipation kernel,
predominate over the scalar field self-interactions. As a result,
under large dissipation,  the
observed evolution for the zero density is quite similar to that of
a linear field, following approximately the
results obtained by replacing, Eq.~(\ref{e2.3.1a}) and (\ref{e2.3.1b}), 
into the expression for the zero  density of a Gaussian theory 
Eq.~(\ref{e2.4.1}). This procedure yields

\begin{equation}
< n(t) > \sim  1/ \left({k_c~ t}\right)^{1/2}.  
\label{e3.2.1}
\end{equation}

Perhaps the most interesting feature of Eq.~(\ref{e3.2.1}) is that it  
constitutes a lower bound on the defect density obtained from 
the classical theory, computed with the same initial field
configuration at the same 
instant in time, for an evolution with any given $\eta$.
This fact remains interesting only for  times that are not too large, 
since  the field evolution under high dissipation quickly
freezes in with  a definite  number of defects (see fig. \ref{fig5}), 
while Eq.~(\ref{e3.2.1}) tends hopelessly to zero densities as time
increases. This discrepancy for large times can be clearly seen in
fig. \ref{fig13}.
 
In contrast to the previous situation, for evolutions under small
external dissipation, the discrepancies between the zero density 
computed from the full classical theory and the linearized theory are
quite dramatic at the time of re-heating.  
The re-heating process is accompanied by the production of 
a large number of zeros and defects that naturally go completely 
unaccounted for by the linear evolution. This is clearly visible in
fig. \ref{fig11}.

The results obtained using the Hartree evolution also fail to 
reproduce those of the full classical evolution but in
somewhat the opposite way.
At re-heating the flow of energy from the infra-red 
to all other modes is very efficient in the Hartree
evolution. When $\eta$ is small  this indeed  results in
essentially the same number of created zeros as in the full
evolution, as can be seen, e.g., in fig. \ref{fig11}. 
The collisionless character of the approximation, however,
precludes the field from ridding itself of high gradient
configurations on small scales. This results in the
fundamental difference between the Hartree and the full
evolution that, in the absence 
of dissipation, the number of zeros decreases in the 
latter but remains approximately constant after creation in the former. 
For short times, of the order of up to $t=20-30$, 
the Hartree approximations therefore yields a number of zeros always
larger than the full classical evolution. For larger times and for reasonably 
dissipated systems ($\eta \geq 0.05$), 
the fact that the Hartree approximation leads to field
configurations with more power on small scales 
than the full evolution implies that it
can be more efficiently dissipated. 
The end result is than after the dissipation makes its effect
noticeable, the zero density resulting from the Hartree evolution
is smaller than that of the full theory. 
For sufficiently large times such densities nevertheless tend to a 
constant value thus freezing in,  as in the case of the
full classical theory, unlike what happens in the case of the 
linear evolution.

For evolutions under stronger dissipation and for relatively short
times, such as can be seen in fig. \ref{fig12}, the discrepancy
between the Hartree and the full classical evolution can be quite
spectacular and the linear result turns out to constitute a much better 
approximation. This is a clear concrete example in which the presence
of a truncated set of interactions actually leads to a much worse
prediction than what could be obtained much more simply from a trivial,
exactly solvable linear approximation.   

\subsection{Correlation Volume Defect Densities, Scaling and the
Kibble Mechanism}
\label{ss3.4}

The evolution of the defect densities and of the field's  correlation length
are interesting on their own right. However, it is clear that they can
constitute two different ways of probing the same qualitative 
structure. 
Defects are associated with sites of space where the
field changes quickly between its two distinct energetically equivalent
minima and constitute therefore regions where the field two-point
correlation  decreases. For this very reason defect densities tell us
statistically how many areas of almost fully correlated field exist
in a given volume, allowing us to probe their mean characteristic
size. This is in turn clearly a measure of the correlation length.

The defect densities per correlation length volume $V_\xi$ constitute
therefore very interesting quantities, that lie at the basis of
fundamental conjectures for defect and domain formation and evolution
such as the Kibble mechanism and  Scaling conjectures.

In figs. \ref{fig14} and \ref{fig15} 
we plot these densities for two different sets of
initial conditions, with small and large initial correlation length,
respectively.

The Kibble mechanism invokes precisely the value of the field's
correlation length in order to predict the defect density produced
at a symmetry breaking phase transition. The correlation length is
clearly a qualitative measure of the size of the volume over which the
field has only small amplitude fluctuations. Defects on the other hand
correspond to field configurations that interpolate between the
energetically indistinguishable minima and should therefore lie at the
boundaries of average correlated patches. At distances larger than the
correlation length the field will be in either one of these
minima. The Kibble mechanism further assumes in what is  usually
designated  the geodesic rule that between such uncorrelated 
regions the continuous
field should take the shortest path over its vacuum manifold. This
results in the simple but powerful prediction that between two correlation
volumes, in one spatial dimension, there will be a defect or not with
probability of 1/2. This is naturally the predicted value for the
correlation volume defect density.

Observing fig. \ref{fig14} and \ref{fig15} it is quite striking 
to notice that regardless of the value of
$\eta$ and of the particular choice of initial conditions, 
the  defect densities obtained at re-heating and thereafter 
are of the order of the prediction  given by the Kibble mechanism 
within a factor of about $10 \%$. 
 
During the stage of re-heating it is still quite apparent that energy
transfers between different scales are large and that arguments based
on local 
energy minimization will be much blurred by large fluctuations
\cite{BER}. Consequently it is  more interesting  to investigate the
behavior of the  defect densities for times sufficiently large that the
system will already have found an energy balance among all scales. 
Fig. \ref{fig16} shows the evolution of the defects densities 
per correlation volume,
for two  relatively close low values of the dissipation $\eta=0.05$
and $\eta=0.01$, and a larger time range.
It is clear that the  correlation volume density of defects tends to a
constant for large times. This is evidence for the  scaling behavior
of the domain wall network, i.e., for the fact that the
number of defects per correlation volume remains constant in
time, approximately once re-heating is complete.

This is an extremely powerful observation as it states that the
{\em statistical} evolution of a domain wall network is characterized by one
single length scale. According to this conjecture the domain wall
network is self-similar after rescaling, 
when in the scaling regime, and its statistical features
can be known at all times given the knowledge of the correlation
length. Fig. \ref{fig16} also shows clearly that, for the same value of
dissipation but different initial conditions the correlation length
defect densities for evolutions under small dissipation result in the
same large time defect densities. This is evidence for the fact that the
 scaling densities are in these
circumstances also independent of the initial conditions, at energies 
above the transition.   

Moreover the  value for the defect density in the scaling regime, 
for the smaller dissipation parameter in fig. \ref{fig16}, is very
close to that predicted by the Kibble mechanism. In contrast, it 
is extremely interesting to notice that for a slightly 
larger value of
$\eta$, the scaling defect density is about $10 \%$ lower. This seems
to indicate that the presence of additional degrees of freedom that
compete with those of the scalar field for its energy, necessarily
leads to scalar field configurations with lower scaling defect
densities.

For evolutions under stronger dissipation the scaling regime is of no
particular interest, however, as it corresponds to the trivial 
situation in which the field is frozen in  and both defect 
densities and correlation length become constant in time.  
Both these remarks may have interesting
realizations as scenarios for defect formation in the early Universe.

\section{Conclusions}
\label{s4}

We presented a detailed study of the non-equilibrium 
dynamics of a classical scalar field theory in a symmetry breaking 
phase transition. In doing so we were able to probe regimes of field
evolution where perturbation theory breaks down and extract
conclusions about the detailed correlation length and topological 
defect density evolution. We showed that after an instantaneous
quench a field develops momentary instabilities responsible for the
growth of its amplitude during which the evolution is approximately
linear, and on a later stage when stability around the
true minimum is found evolves, in a period of re-heating very similar
to that of inflationary scenarios. During this later stage the
evolution proceeds   in a strongly non-linear 
fashion so as  to redistribute energy among all scales. 
We showed the effect of external dissipation in the evolution and
established evidence for the independence of 
defect densities per correlation length on the choice of initial
conditions as well as their approach to a scaling regime for large
times. In passing we confronted the predictions of the Kibble mechanism
to our results 
and discussed the effect of the external dissipation on the asymptotic
defect densities. We have also shown the comparison between the zero
densities given by the full classical evolution and the predictions from
the linearized theory and the Hartree approximations thus clarifying
when these approximate schemes are valid.

Finally we believe that the  present work raises  many 
new important questions about the evolution of
relativistic fields away from thermal equilibrium. 
We are presently investigating the effect of an expanding background
on topological defect production and evolution, in higher spatial
dimensions, and studying the details of the energy redistribution during
the phase transition. Both these subjects have
important cosmological applications to the role of topological defects
as the seeding mechanism  for the formation of structure
in the Universe and the theory of re-heating after a period of 
inflationary expansion.

\section*{Acknowledgements}

It is a great pleasure for us to thank
Salman Habib, Wojciech Zurek and Marcelo Gleiser 
for extremely useful suggestions.
We also thank Ed Copeland, Ray Rivers and Graham Vincent
for discussions and
Alisdair Gill for comments on the manuscript.  
NDA's and LMAB's research was supported by J.N.I.C.T.- {\it 
Programa Praxis XXI},
under contracts BD/2794/93-RM  and BD/2243/92-RM, respectively. 
This work was  supported in part by the European Commission under 
the Human Capital and Mobility programme, contract no. CHRX-CT94-0423.
We  thank The Fujitsu/Imperial College Centre for Parallel Computing 
for generous allocation of resources.

\begin{figure}
\centerline{\psfig{file=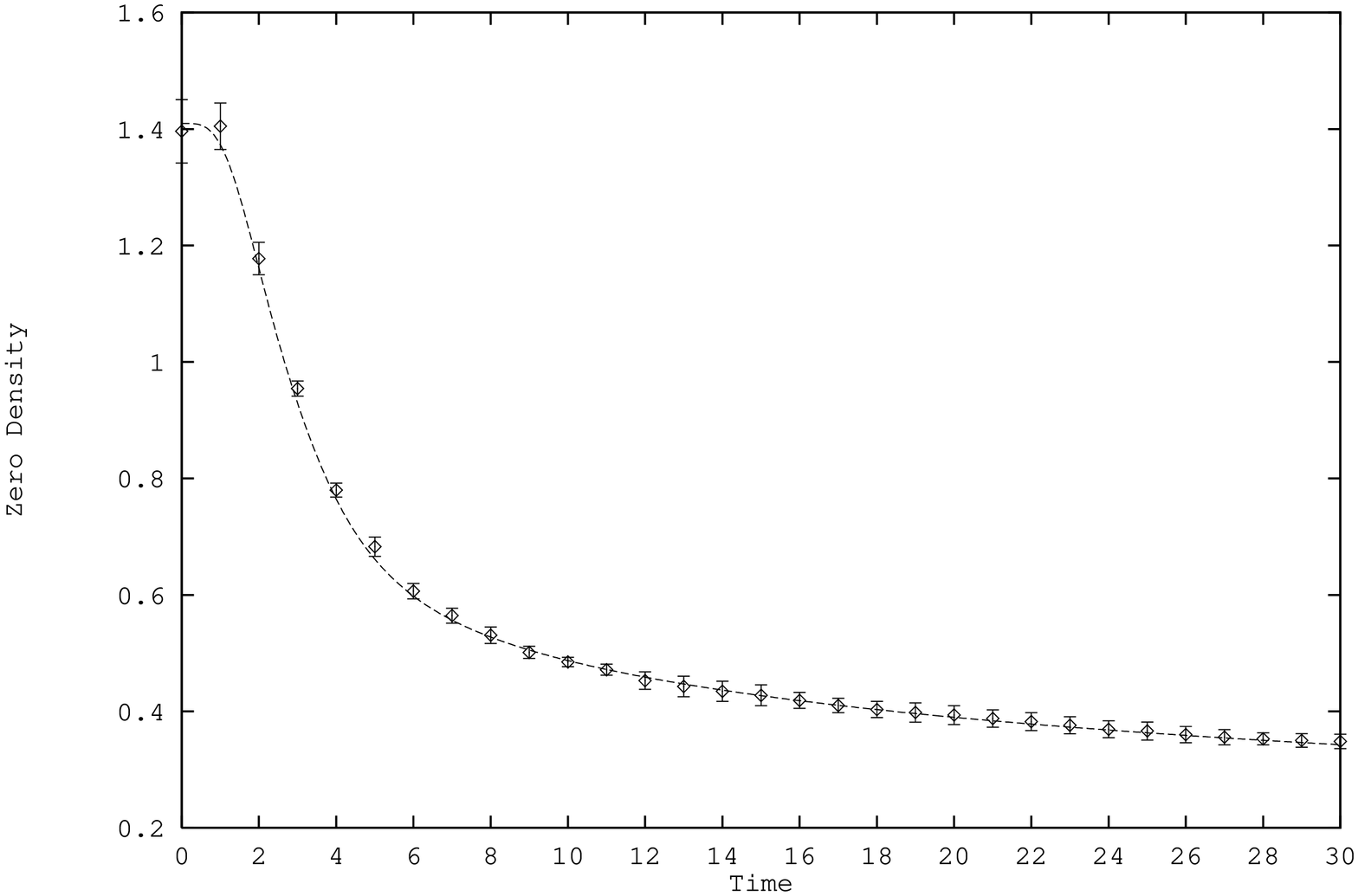,width=2in,angle=270}}
\caption{Exact Linear evolution (solid line) and numerical evolution
(data points) for the zero density in the free case. 
The error bars denote the 
standard deviation from the mean, computed over our ensemble 
of 128 realizations.}
\label{fig1}
\end{figure}

\begin{figure}
\centerline{\psfig{file=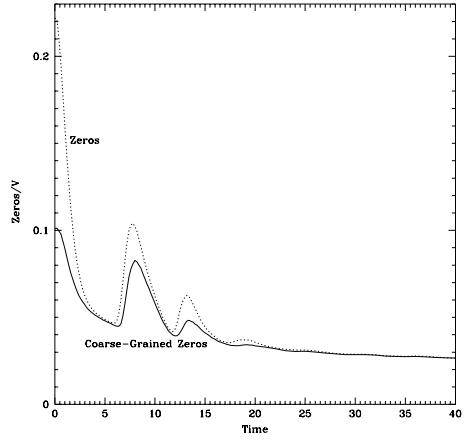,width=2.5in}}
\caption{Evolution of the total zero density and that in the coarse
grained field on the scale of the width of a defect for $\eta=0.05$
and initial conditions with  $M=0.1$, $T=0.005$. For late times the
coarse-grained zeros  can be identified as topological defects.}
\label{fig2}
\end{figure}

\begin{figure}
\centerline{\psfig{file=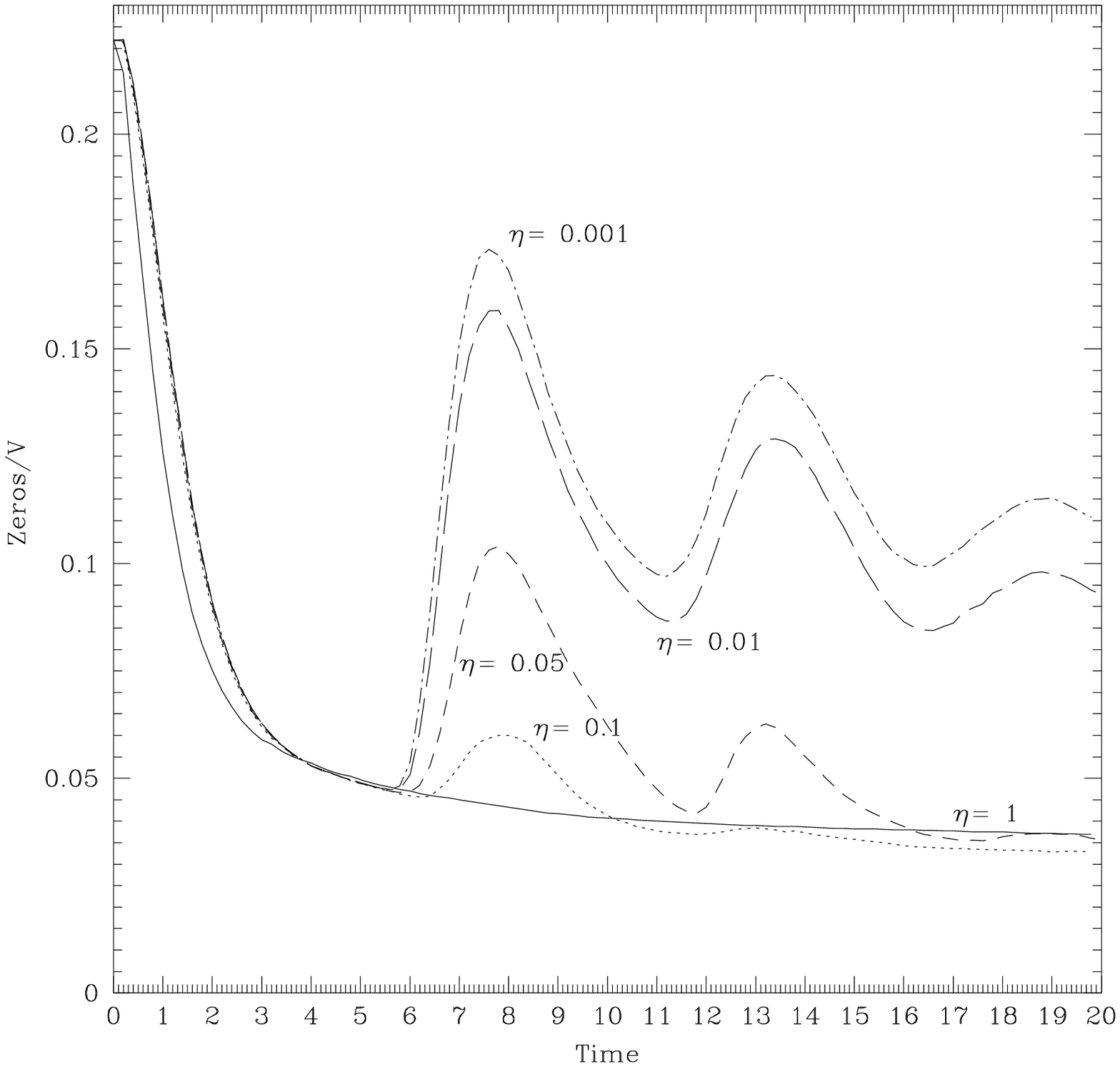,width=2.5in}}
\caption{Evolution of the zero density for different values of $\eta$
and initial conditions with $M=0.1$ and $T=0.005$.}
\label{fig3}
\end{figure}

\begin{figure}
\centerline{\psfig{file=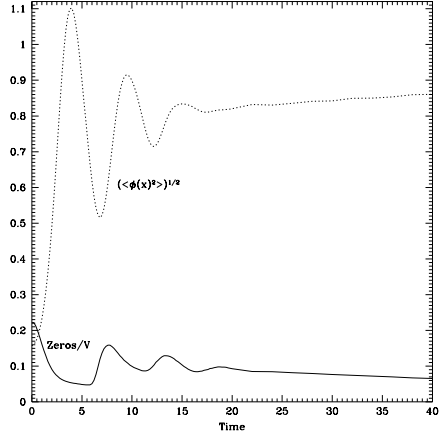,width=2.5in}}
\caption{The production of zeros at re-heating and the mean field 
evolution. It is clear that zero production proceeds by bursts.}
\label{fig4}
\end{figure} 

\begin{figure}
\centerline{\psfig{file=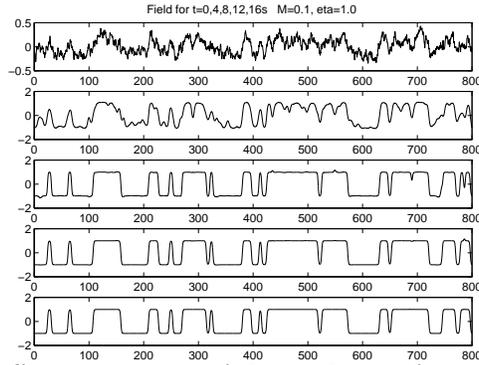,width=2.5in}}
\caption{Field profiles at different instants of the evolution for  
$\eta=1$ and initial conditions with $M=0.1$ and $T=0.005$. It is
evident that the field configuration is frozen in for the larger times.}
\label{fig5}
\end{figure}

\begin{figure}
\centerline{\psfig{file=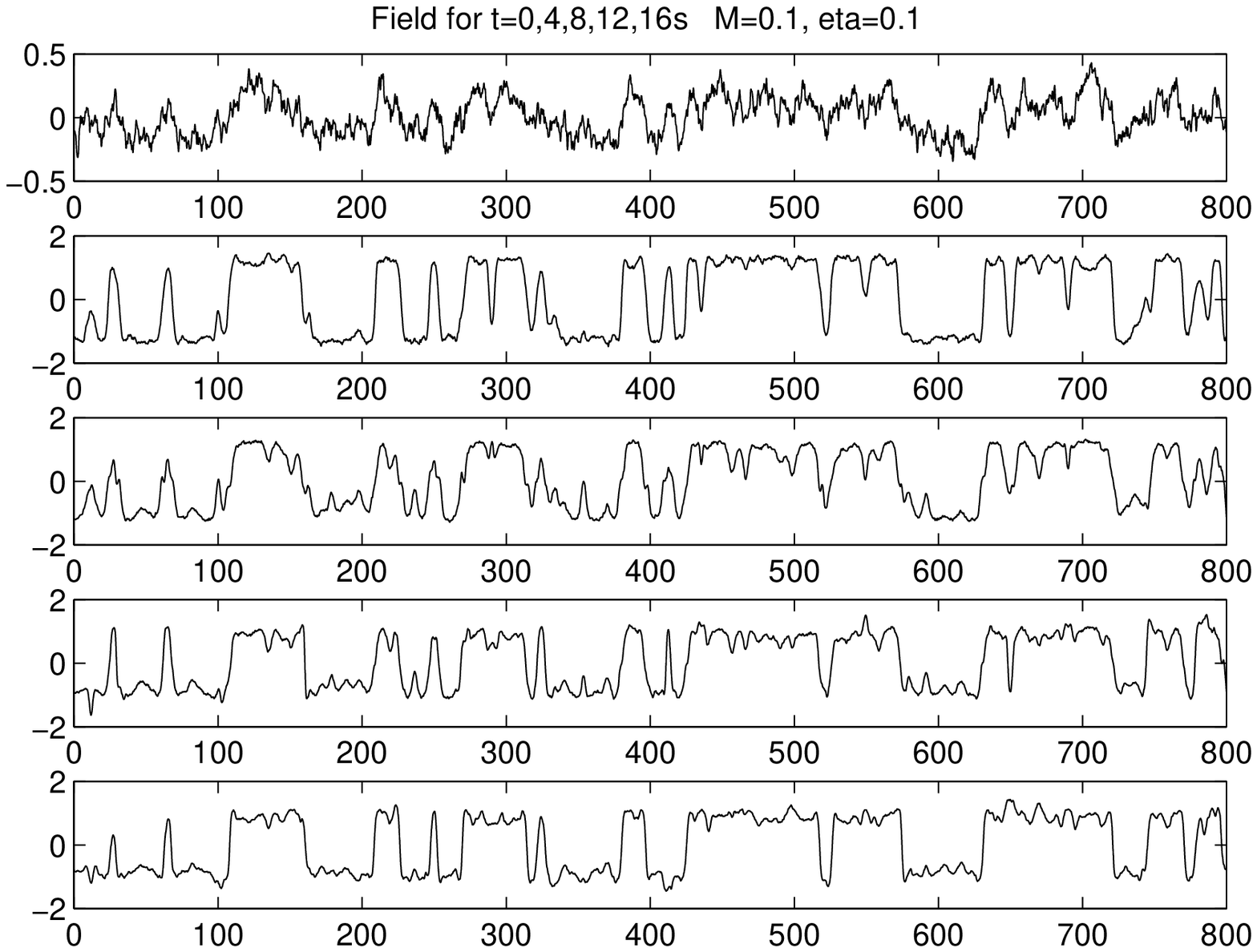,width=2.5in}}
\caption{Field profiles at different instants of the evolution for  
$\eta=0.1$ and initial conditions with $M=0.1$ and $T=0.005$.}
\label{fig6}
\end{figure}

\begin{figure}
\centerline{\psfig{file=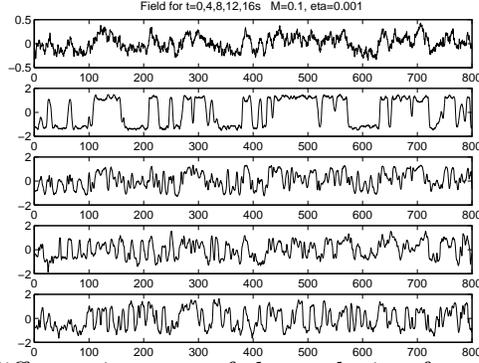,width=2.5in}}
\caption{Field profiles at different instants of the evolution for  
$\eta=0.001$ and initial conditions with $M=0.1$ and $T=0.005$. The
effects of re-heating are clear from the comparison of the second
frame to the third.}
\label{fig7}
\end{figure}

\begin{figure}
\centerline{\psfig{file=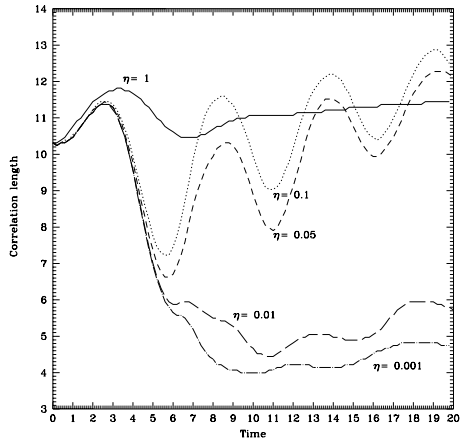,width=2.5in}}
\caption{Time Evolution of the Correlation Length for $M=0.1$, $T=0.005$ and 
several values of the dissipation $\eta$.}
\label{fig8}
\end{figure}

\begin{figure}
\centerline{\psfig{file=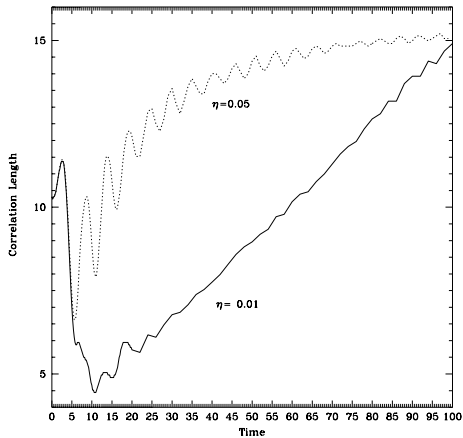,width=2.5in}}
\caption{Long Time Evolution of the Correlation Length 
for initial conditions with $M=0.1$, $T=0.005$  
$\eta=0.01$ and $\eta=0.05$. For the smaller of these values the
correlation length evolves linearly with coefficient $0.1$ whereas for
the larger dissipation it displays approximately diffusive behavior.}
\label{fig47}
\end{figure}

\begin{figure}
\centerline{\psfig{file=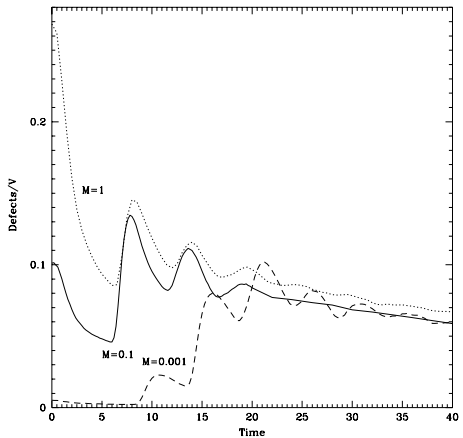,width=2.5in}}
\caption{Defect density evolution for three different initial
conditions, with $M=1,T=0.02$, $M=0.1, T=0.005$ and  
$M=0.001,T=1.10^{-5}$.}
\label{fig9}
\end{figure} 

\begin{figure}
\centerline{\psfig{file=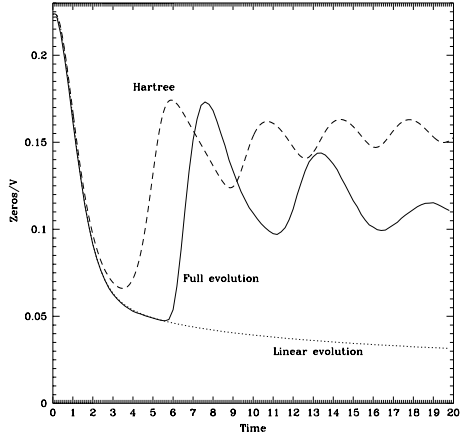,width=2.5in}}
\caption{The zero density given by the full classical, linearized
and Hartree evolutions for $\eta=0.001$ and initial conditions with
$M=0.1$ and $T=0.005$.}
\label{fig11}
\end{figure}

\begin{figure}
\centerline{\psfig{file=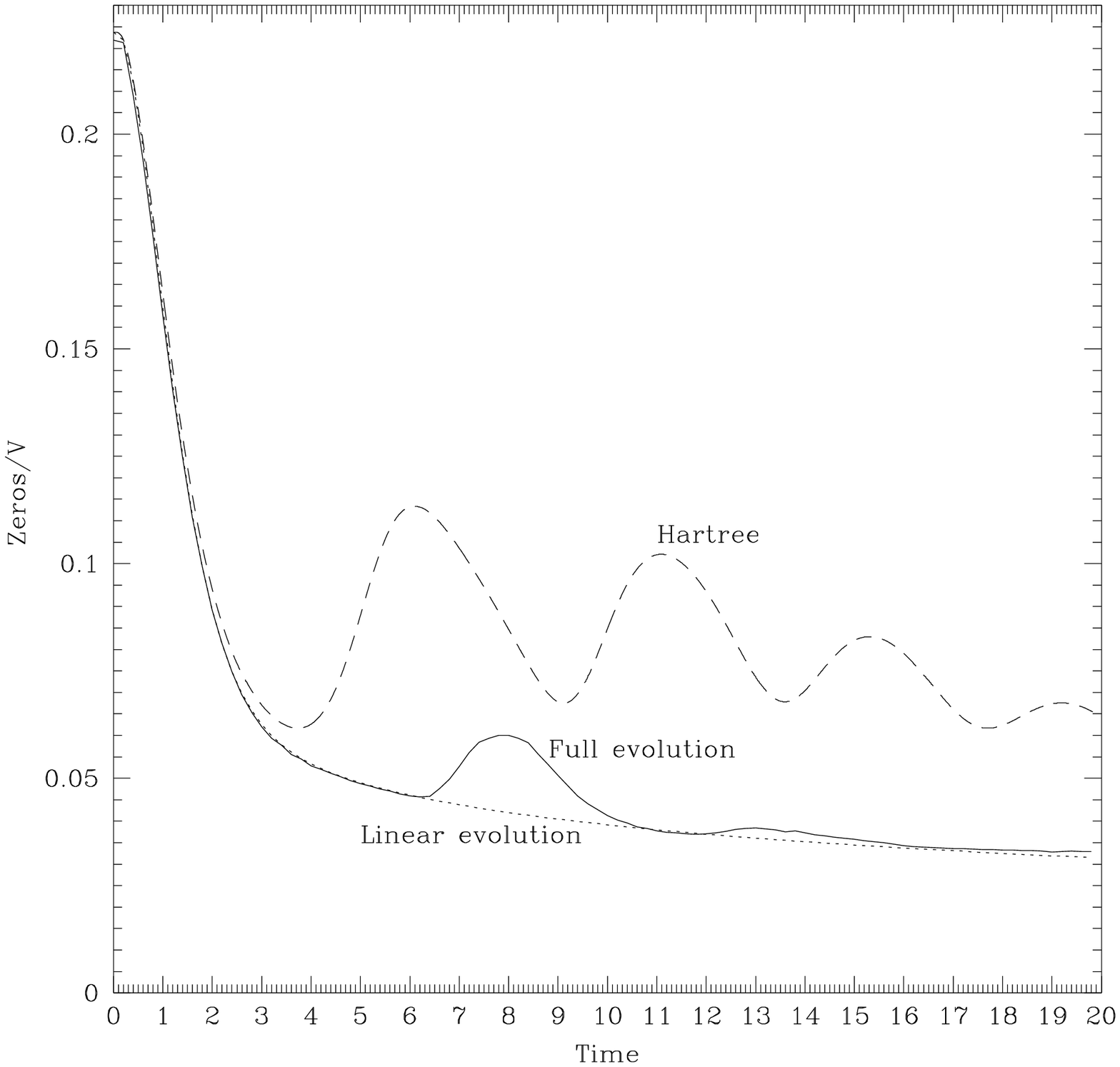,width=2.5in}}
\caption{The density of zeros given by the full classical, linearized
and Hartree evolutions for $\eta=0.1$ and initial conditions with
$M=0.1$ and $T=0.005$.}
\label{fig12}
\end{figure}

\begin{figure}
\centerline{\psfig{file=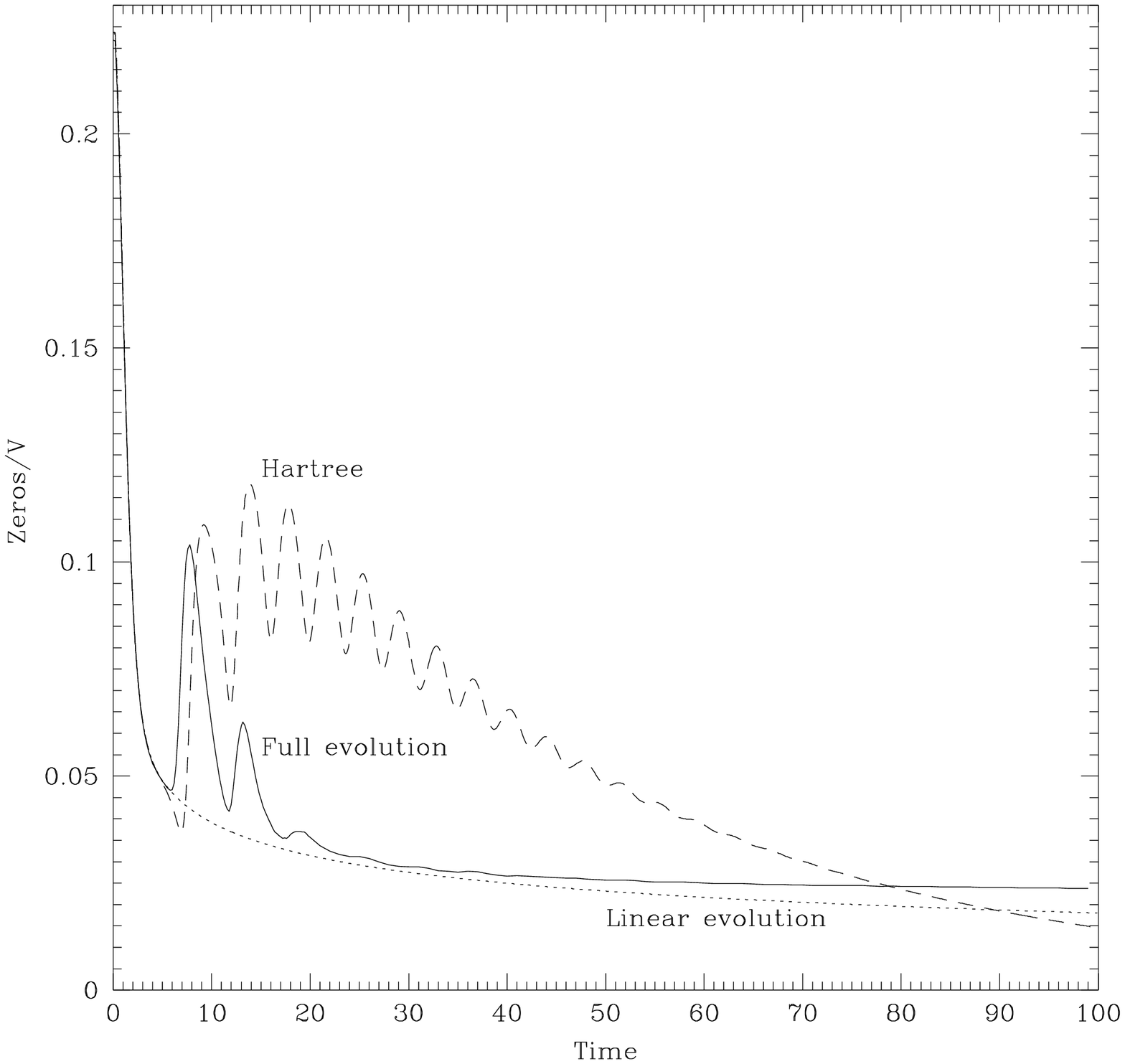,width=2.5in}}
\caption{The density of zeros given by the full classical, linearized
and Hartree evolutions for $\eta=0.05$ and initial conditions with
$M=0.1$ and $T=005$.}
\label{fig13}
\end{figure}

\begin{figure}
\centerline{\psfig{file=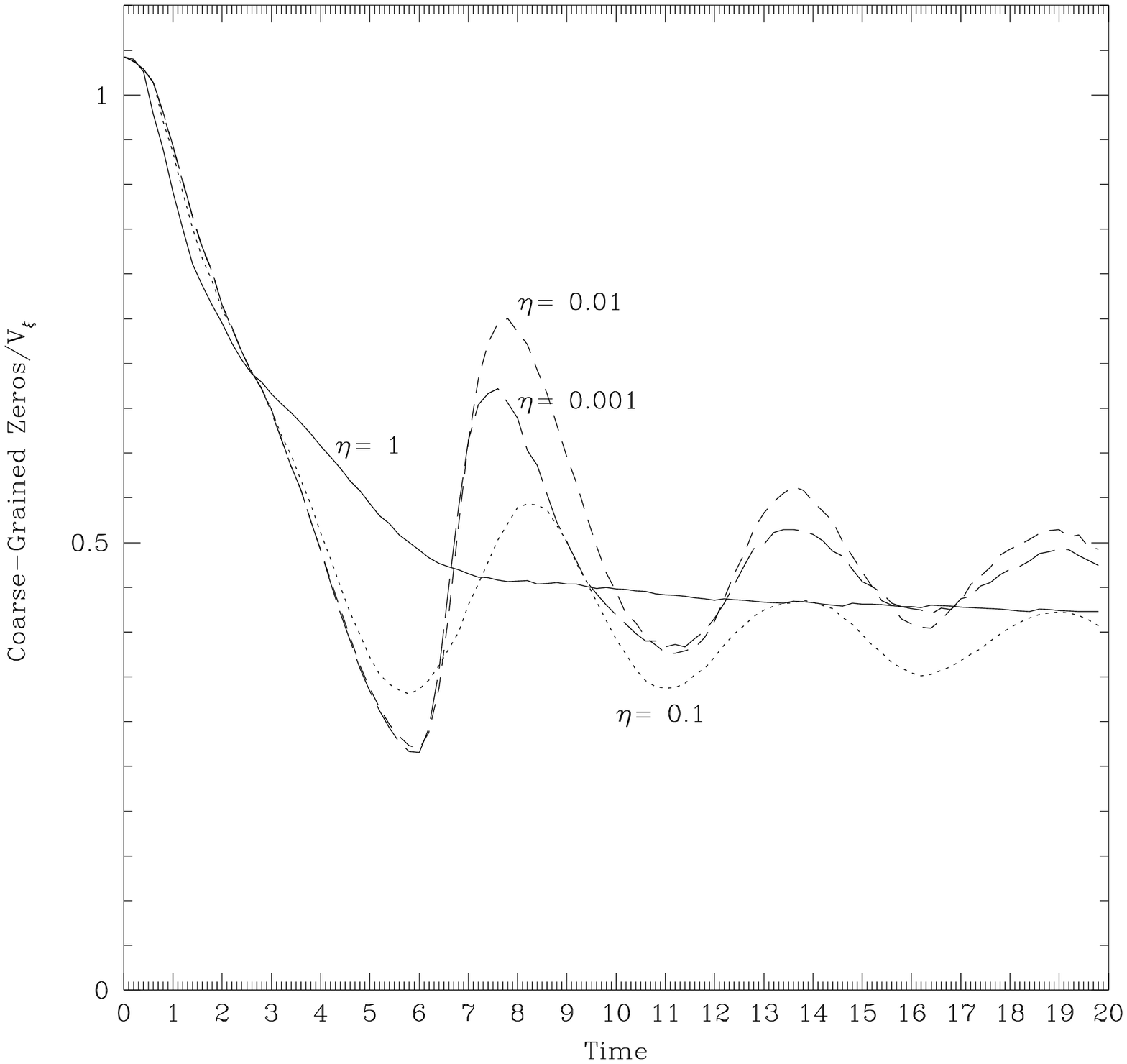,width=2.5in}}
\caption{The evolution of the defects densities per correlation volume
for several values of $\eta$, for initial conditions with $M = 0.1$
and $T=0.005$.}
\label{fig14}
\end{figure}

\begin{figure}
\centerline{\psfig{file=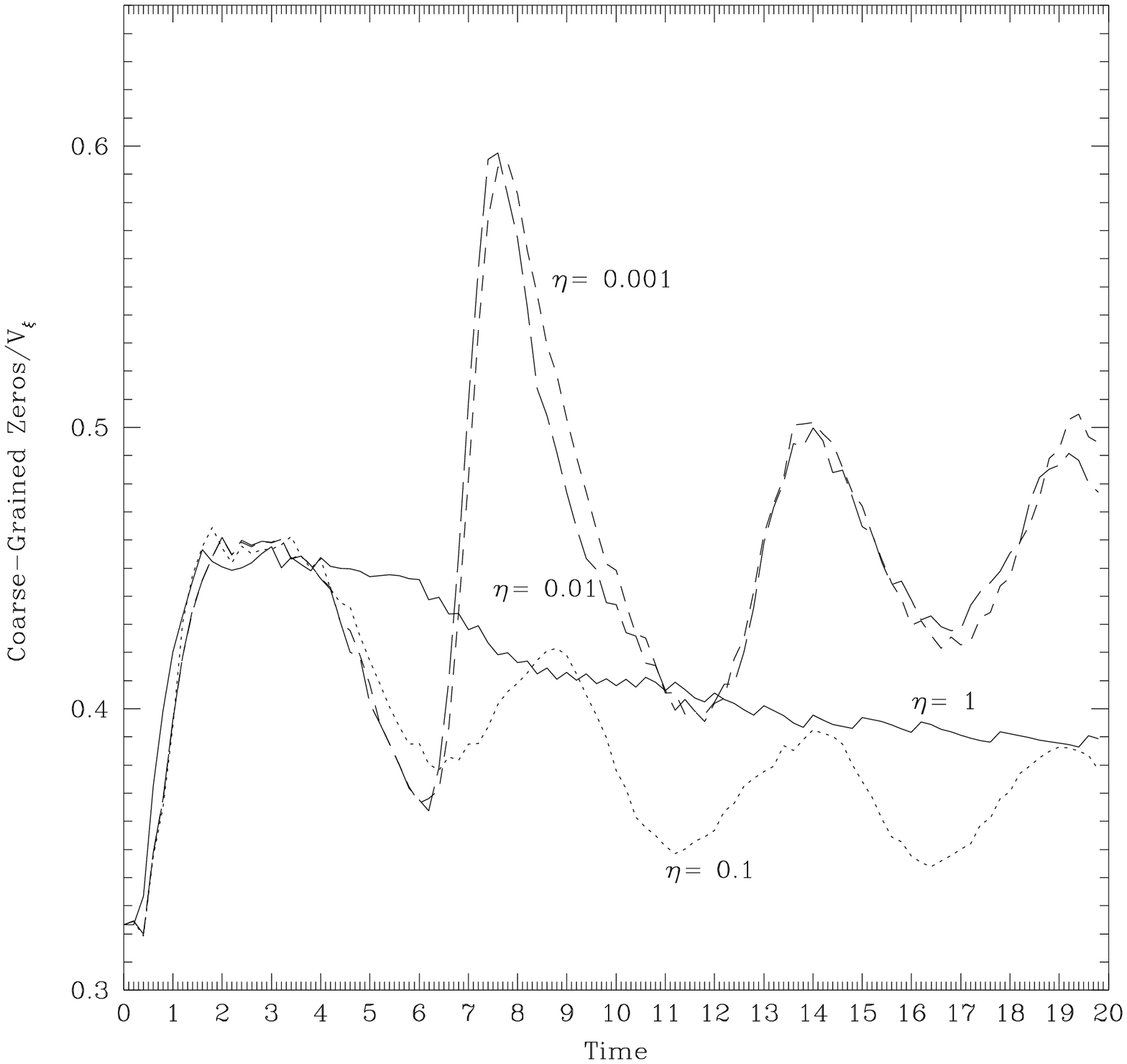,width=2.5in}}
\caption{The evolution of the defects densities per correlation volume
for several values of $\eta$, for initial conditions with  
$M = 1$ and $T=0.02$.}
\label{fig15}
\end{figure}

\begin{figure}
\centerline{\psfig{file=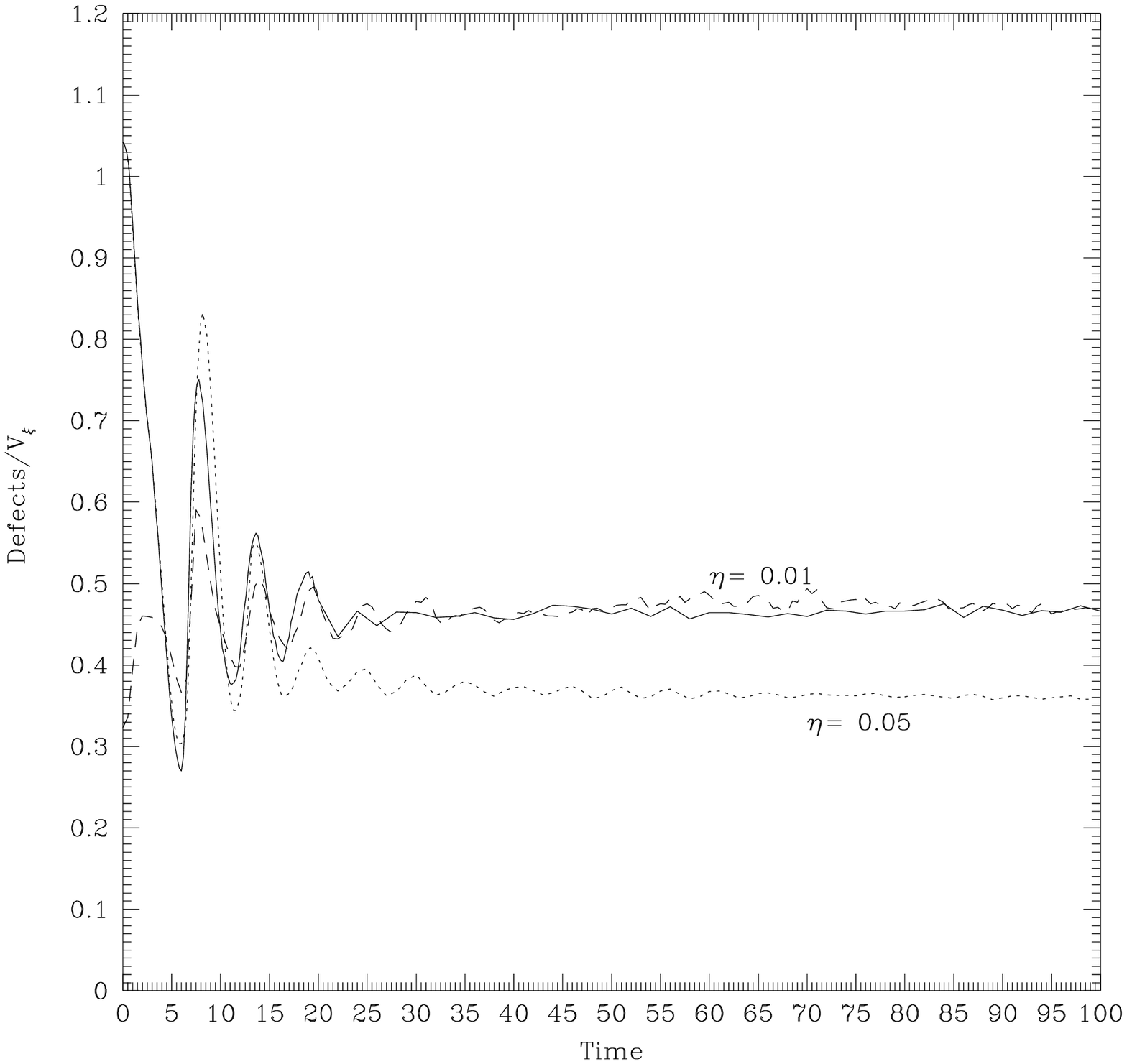,width=2.5in}}
\caption{The evolution of the defects densities per correlation volume
for two values of $\eta$. For $\eta =0.01$ the dashed line corresponds to
initial conditions with $M = 1$ and the solid line with $M=0.1$ 
The scaling regime for large times is apparent as well as the
corresponding defect density independence on the choice of initial
conditions, for $\eta=0.01$.}
\label{fig16}
\end{figure}


\begin{thebibliography}{99} 
\bibitem{Kib} T. W. B. Kibble, J. Phys. A {\bf 9}, 1387 (1976)
\bibitem{Book} For a comprehensive review see  A. Vilenkin and 
 E. P. S. Shellard, {\it Cosmic Strings and
 other Topological Defects} (Cambridge University Press, Cambridge, England,
 1994).
\bibitem{ZurekReview} W.H. Zurek, {\it Los Alamos
preprint} LA--UR--95--2269, Phys. Rep., in press; Acta Physica Polonica B
{\bf 24}, 1301--1311 (1993).
\bibitem{Helium4} P. C. Hendy, N. S. Lawson, R. A. M. Lee, P. V. E.
 McLintock and C. D. H. Williams, Nature {\bf 368}, 315 (1994).
\bibitem{Helium3}  V.M.H. Ruutu et al., Imperial College Pre-print 
IMPERIAL/TP/95--96/17, accepted for publication in Nature.
\bibitem{liqcrys} I. Chuang, R. Durrer, N. Turok and B. Yurke, Science
 {\bf 251}, 1336 (1991) ; M.J. Bowick, L. Chander, E. A. Schiff
 and A. M. Srivastava, Science {\bf 263}, 943 (1994) 943.  
\bibitem{VV} T. Vachaspati and A. Vilenkin, Phys. Rev. D {\bf 31}, 3052 (1985)
\bibitem{Struc}Ya. B. Zeldovich, Mon. Not. Astron. Soc. {\bf 192}, 663
(1980); A. Vilenkin, Phys. Rev. Lett.  {\bf 46}, 1169 (1981); {\bf
46}, 1496(E) (1981). 
\bibitem{quench} D. Boyanovsky, D.-S. Lee, A. Singh, Phys. Rev. D {\bf 48},
 800 (1993)
\bibitem{non-eq} D. Boyanovsky and H. J. de Vega, Phys. Rev. D {\bf 47}, 2343
  (1993); H. J. de Vega and R. Holman , Phys. Rev. D {\bf 49},
  2769 (1994);  D. Boyanovsky,
H. J. de Vega, D.-S. Lee and A. Singh, Phys. Rev. D {\bf 51}, 4419 (1995).
\bibitem{LA} F. Cooper, S. Habib, Y. Kluger, E. Mottola, J. P. Paz and
 P. R. Anderson, Phys. Rev. D {\bf 50}  2848 (1994).
\bibitem{Ray}  A. J. Gill and R. J. Rivers, Phys. Rev. D {\bf 51},6949
(1995); G. Karra and R. J. Rivers,  Imperial College preprint 
Imperial/TP/95-96/28 and hep-ph/9603413. 
\bibitem{Parisi} G. Parisi, {\it Statistical Field Theory} (Addison-Wesley
 Publishing  Company, Inc. 1988)  
\bibitem{MW} N. D. Mermin and H. Wagner, Phys. Rev. Lett. {\bf 22}, 1133
(1966) 
\bibitem{ABY} N. D. Antunes, L. M. A. Bettencourt and A. Yates, in preparation.
\bibitem{Kabib} S. Habib and H. E. Kandrup, Phys. Rev. D {\bf 46}, 5303
(1992) 
\bibitem{Halperin} B. I. Halperin, {\it Statistical Mechanics of Topological
 Defects}, published in {\it Physics of Defect}, proceedings of Les Houches,
 Session XXXV 1980 NATO ASI, editors Balian, Kl\'{e}man and Poirier 
 (North-Holand Press, 1981), 816 (1981). For a more comprehensive
derivation see F. Liu and G. F. Mazenko, Phys. D Rev. {\bf 46}, 5963 (1992). 

\bibitem{Habib} S. Habib in Proceedings of NEEDS'94, Los Alamos,
 (September 1994)
\bibitem{HF} S. J. Chang, Phys. Rev. D {\bf 12}. 1071 (1975)
\bibitem{BER} L. M. A. Bettencourt, T. S. Evans and R. J. Rivers, 
Phys.Rev. D {\bf 53}, 668, (1996).
 
\end{thebibliography}
\end{document}